\documentclass[12pt,a4paper]{article} 
\pdfoutput=1
\usepackage{style}

\usepackage{color,graphicx,amsmath,amssymb,lipsum}

\usepackage[normalem]{ulem}

\usepackage[utf8]{inputenc}
\usepackage[T1]{fontenc}
\usepackage{mathtools}
\usepackage{bm}

\usepackage{multirow}

\usepackage{ulem}
\usepackage{diagbox}

\usepackage[dvipsnames]{xcolor}
\newcommand{\rdrag}{r_{\rm drag}}
\newcommand{\kms}{{\rm \,\,km\,s^{-1}Mpc^{-1}}}

\newcommand{\eV}{\,\text{eV}}

\renewcommand\thesection{\arabic{section}}

\newcommand{\be}{\begin{equation}}
\newcommand{\ee}{\end{equation}}
\newcommand{\beqa}{\begin{eqnarray}}
\newcommand{\eeqa}{\end{eqnarray}}
\newcommand{\bseq}{\begin{subequations}}
\newcommand{\eseq}{\end{subequations}}

\renewcommand\O{\Omega}

\newcommand\s{\sigma}

\renewcommand\L{\Lambda}

\renewcommand{\ln}{\mathop{\rm ln}\nolimits}

\def\Mnu{\sum m_\nu}

\def\beqn#1{\begin{equation}\label{#1}}
\def\eeqn{\end{equation}}

\def\beqa#1{\begin{eqnarray}\label{#1}}
\def\eeqa{\end{eqnarray}}

\def\Z2{$\mathcal{Z_2}$}

\newcommand {\ignore}[1]{}

\title{Relaxing DESI DR2 BAO Constraints\\on $\Mnu$ with Planck and SPT-3G 2018\\in the Context of SPT D1

}

\affiliation[b]{Institute for Nuclear Research of the Russian Academy of Sciences, \\ 
60th October Anniversary Prospect, 7a, 117312
Moscow, Russia}
\affiliation[c]{Moscow Institute of Physics and Technology,\\
	Institutsky lane 9, Dolgoprudny, Moscow region, 141700, Russia}
\affiliation[d]{Department of Particle Physics and Cosmology, Physics Faculty, M.V. Lomonosov Moscow State University, \\ Vorobjevy Gory, 119991 Moscow, Russia}

\author[b,c]{Dmitry Gorbunov\footnote{\texttt{gorby@ms2.inr.ac.ru}}}

\author[b]{, Nikita Nedelko\footnote{\texttt{nedelko@inr.ru}}}

\begin{document}

\abstract{
We present constraints on the sum of neutrino masses $\Mnu$ from a dataset incorporating the full SPT-3G 2018 TT/TE/EE+lensing spectra together with Planck PR4 lensing and low-$\ell$ parts of the Planck PR3 spectra. Using it as a baseline for the DESI DR2 BAO measurements, we arrive at a $95\%$ upper limit of $\Mnu < 0.11 \eV$, relaxing the tension between $\rm \Lambda$CDM and lower bounds on $\Mnu$ based on neutrino oscillation experiments. When including DES Y1 weak lensing information and the Pantheon+ SNIa catalog, the limit is further loosened to $\Mnu<0.138 \eV$ with a slight preference for $\Mnu>0$. On contrast, replacing SPT-3G 2018 primary CMB and lensing spectra with ones from the SPT-3G 2019-2020 (D1) release tightens the overall constraint to $<0.082 \eV$ and pushes the $\Mnu$ posterior mode value to zero, indicating a preference for quasi-negative neutrino masses in line with the D1 analysis. This is a curious shift within SPT-3G measurements of the same field taken in 2018 and in 2019-2020 and processed with different analysis pipelines.

}

\begin{flushright}
\end{flushright}

\maketitle
\flushbottom

\section{Introduction}
\label{sec:intro}

Ever since the discovery of neutrino oscillations, finding the scale of neutrino masses has been one of the biggest ongoing quests in fundamental physics. As neutrino oscillations only depend on the differences of squared neutrino masses, few ground-based experiments can limit the neutrino masses from above, with the KATRIN experiment putting an upper limit of $0.45 \eV$ on the mass of the electron neutrino~\cite{KATRIN:2024cdt}, which may be translated on the mass of the lightest eigenstate. On the other hand, the precise measurement of $\Delta m^2_{ij}$ provides a lower limit on the sum of neutrino masses by assuming that the lightest state is massless, with the lower bound currently standing at $\approx 0.06 \eV$ for the normal neutrino mass hierarchy and $\approx 0.10 \eV$ for the inverse  hierarchy~\cite{Esteban:2018azc}.  

Cosmological observations can complete the puzzle. The mass of neutrinos changes the evolution of the Universe and in particular the growth of its large-scale structure (LSS)~\cite{Racco:2024lbu}. However, till now no signature of neutrino mass impacts on the cosmology has been observed. This allowed cosmologists to place upper limits on the sum of neutrino masses $\sum m_\nu$ from a combined analysis of the CMB measurements and galaxy-mapping surveys. 

Recently, first and second data releases of the DESI experiment have become available~\cite{DESI:2024lzq,DESI:2024uvr,DESI:2025zpo}, providing new precise measurements of the LSS. So far, these measurements have been combined with the full-sky CMB maps produced by the Planck space telescope~\cite{Planck:2018vyg,Tristram:2023haj}, and the more precise (at small angular scales/large multipoles $\ell$) partial maps from the Atacama Cosmology Telescope (ACT)~\cite{ACT:2023dou}, resulting in strict upper bounds of $\sum m_\nu <0.07 \eV$~\cite{DESI:2024hhd,DESI:2024mwx,DESI:2025zgx} (at $95\%$ confidence level), putting them in tension with the 3-flavor picture of neutrino oscillations, especially in the case of inverse mass hierarchy. In this context we turn to the other investigator of CMB properties, the SPT-3G project~\cite{SPT-3G:2014dbx}, which produces high-resolution maps of a relatively small part of the sky, in conjunction with low-$\ell$ and lensing data from Planck, to construct two competitive "SPT-forward" CMB datasets based on the results of 2018 and 2019-2020 SPT-3G observational seasons. Using it as a baseline and incorporating information from DESI DR2 BAO~\cite{DESI:2025zgx}, DES Y1 weak lensing~\cite{DES:2017myr} and the Pantheon+ SN catalog~\cite{Scolnic:2021amr}, we present a set of constraints on $\Mnu$. We find that while SPT-3G 2018 data is generally able to reconcile DESI DR2 BAO measurements with massive neutrinos when Pantheon+ Type Ia supernovae are included, SPT-3G 2019-2020 closes this window.

\section{The pipeline}
\label{sec:param}

The first baseline dataset, referred to as '\texttt{CMB}', is built around SPT-3G Main field observations from the 2018 season. It includes:
\begin{itemize}
    \item the full SPT-3G 2018 $TT/TE/EE$ CMB maps~\cite{SPT-3G:2022hvq} as implemented in the \texttt{candl} package~\cite{Balkenhol:2024sbv} (covering $750\leq \ell\leq 3000$ for $TT$ and $300\leq \ell\leq 3000$ for $TE/EE$)
    \item $\ell<30$ Planck PR3 $TT$ maps in the \texttt{Commander} realization~\cite{Planck:2019nip}
    \item $\ell<30$ Planck PR3 $EE$ maps in the \texttt{SRoll2} realization~\cite{Delouis:2019bub}
    \item $30<\ell<756$ Planck PR3 $TT$ maps in the \texttt{Plik} realization~\cite{Planck:2019nip}
    \item CMB lensing potential spectra from  SPT-3G 2018~\cite{SPT:2023jql} 
    \item CMB lensing potential spectra from Planck PR4~\cite{Carron:2022eyg}
\end{itemize}
We choose not to include the Planck high-$\ell$ TE/EE data because the SPT-3G TE and EE spectra extend down to $\ell = 300$, albeit with less constraining power.

The second baseline dataset, '\texttt{CMB-D1}', replaces the 2018 SPT-3G measurements with those from the SPT-3G D1 analysis covering the same field in the seasons of 2019 and 2020:
\begin{itemize}
    \item the full SPT-3G D1 (2019-2020) $TT/TE/EE$ CMB maps~\cite{SPT-3G:2025bzu} as implemented in the \texttt{candl} package~\cite{Balkenhol:2024sbv} (covering $400\leq \ell\leq 3000$ for $TT$ and $400\leq \ell\leq 4000$ for $TE/EE$)
    \item CMB lensing potential spectra obtained from SPT-3G 2019-2020 EE measurements using the MUSE pipeline~\cite{SPT-3G:2024atg} 
    \item The same Planck PR3 primary spectra and PR4 lensing spectra as in $\texttt{CMB}$
\end{itemize}
We keep our cut of the Planck high-$\ell$ spectra unchanged to focus on the difference between SPT-3G 2018 and 2019-2020 data. 

Additionally, we use the following low- and medium-redshift observations to complement CMB data:
\begin{itemize}
\item the full set of DESI DR2 BAO measurements~\cite{DESI:2025zgx}
    \item weak lensing measurements from DES Y1~\cite{DES:2017myr}
    \item the full uncalibrated Pantheon+ catalog of Type Ia supernovae~\cite{Scolnic:2021amr}
\end{itemize}

We produce MCMC chains using the standard Metropolis-Hastings algorithm from the \texttt{Cobaya} suite~\cite{Torrado:2020dgo} interfaced with the \texttt{CLASS} Einstein--Boltzmann solver~\cite{Diego_Blas_2011}. For consistency, we discard the first $20\%$ of each chain during analysis. Plots and parameter estimates are produced using the \texttt{getdist} package~\cite{Lewis:2019xzd}. 

In the $\Lambda$CDM model we vary the following set of cosmological parameters: ($\omega_{cdm}$, $\omega_b$, $H_0$, $\ln(10^{10}A_s)$, $n_s$, $\tau$) where 
$\omega_{cdm}\equiv \Omega_{cdm} (\frac{H_0}{100\kms})^2$, $\omega_b\equiv\Omega_b (\frac{H_0}{100\kms})^2$. Here $\Omega_{cdm}$ and $\Omega_{b}$ refer to the present fractions of cold dark matter and baryons in the energy density of the Universe, $H_0$ is the Hubble constant, $A_s$ and $n_s-1$ stand for the amplitude and spectral index of the matter perturbations, and $\tau$ is the reionization optical depth. 
In $\Lambda$CDM we assume normal neutrino hierarchy with the total active neutrino mass $\sum m_\nu=0.06\eV$. 
In the $\rm \Lambda$CDM+$\sum m_\nu$ model we instead vary $\Mnu$ as a free parameter, approximating the neutrino sector with three degenerate massive states to improve the evaluation speed of \texttt{CLASS} and not making any assumptions about mass hierarchy. Throughout the paper $\sum m_\nu$ is in units of $\eV$ and $H_0$ is in units of $\kms$.




\section{$\Lambda$CDM constraints}
\label{sec:LCDM}
It has already been shown~\cite{Chudaykin:2022rnl} that the SPT-3G 2018 TE/EE dataset is fully compatible with Planck PR3 TT spectra with $\ell_{max}$ of up to 1000. Similarly, the full SPT-3G 2018 TT/TE/EE dataset used in this work doesn't exhibit any tensions with our choice of Planck TT ($\ell<756$) or with the \texttt{SRoll2} polarization spectra, with only minor parameter shifts upon the addition of Planck data. Table\,\ref{tab:tLCDM1} details the constraints produced by incrementally adding Planck data and lensing spectra to SPT-3G. Without CMB lensing we achieve slightly weaker constraints on $\rm \Lambda$CDM parameters compared to Planck PR3 TT/TE/EE~\cite{Planck:2018vyg} (except for $A_s$ and $\tau$, the $1\sigma$ errors on which are on par with those produced by Planck PR4~\cite{Tristram:2023haj} as expected from including low-$\ell$ Planck data).
The SPT-3G 2018 lensing potential spectra only marginally tighten the $\L$CDM constraints while the further addition of Planck PR4 lensing data considerably tightens constraints on $H_0$, $\Omega_m$, and $\sigma_8$ in particular at the expense of $\sim1\sigma$ shifts in these parameters. 
With each new data set added to the analysis, the error bars in the inferred values of the cosmological parameters shrink without any abnormal shifts of mean values. Thus, one may conclude that the CMB data sets are fully consistent. 

\begin{table}
    \small
	\centering
	\begin{tabular} {|c||c|c|c|c|c|}
		\hline
		& \multicolumn{5}{c|}{$\rm \Lambda$CDM} \\
		\hline
		\hline
		\multirow{2}{*}{\!\!\! Parameter\!} &  SPT-3G & $\rm +Planck\,lowT$ & +Planck \,TT & +SPT-3G & +Planck 
		\\
		& TT/TE/EE & $\rm +Sroll2$ & $\rm (\ell<756)$ & $\rm Lensing$ & $\rm PR4~Lensing$ \\
		\hline
		$100\,\omega_b$ & $2.223\pm0.031$ &
		$2.214\pm0.030$ &
		$2.231\pm0.020$ &
        $2.231\pm0.020$ &
		$2.227\pm0.020$  \\ 
		$10\,\omega_{cdm}$ & $1.166\pm0.038$ &
		$1.153\pm0.037$ &
		$1.163\pm0.022$ &
        $1.167\pm0.020$ &
		$1.180\pm0.014$ \\ 
		$H_0$ & $68.29\pm1.51$ &
		$68.68\pm1.52$ &
		$68.54\pm0.96$ &
        $68.36\pm0.88$ &
		$67.80\pm0.65$ \\
		$\tau$ & $0.053\pm0.007$ &
		$0.058\pm0.006$ &
		$0.057\pm0.006$ &
        $0.057\pm0.006$ &
		$0.057\pm0.006$ \\ 
		$\!{\rm ln}(10^{10} A_s)\!$ & $3.034\pm0.020$ &
		$3.034\pm0.015$ &
		$3.037\pm0.013$ &
        $3.040\pm0.012$ &
		$3.045\pm0.012$ \\ 
		$n_s$ & $0.972\pm0.016$ &
		$0.983\pm0.011$ &
		$0.971\pm0.007$ &
        $0.970\pm0.007$ &
		$0.968\pm0.005$ \\ 
		\hline   
		$\Omega_m$ & $0.300\pm0.021$ &
		$0.294\pm0.021$ &
		$0.297\pm0.013$ &
        $0.299\pm0.012$ &
		$0.307\pm0.009$ \\ 
		$\sigma_8$ & $0.797\pm0.015$ &
		$0.796\pm0.014$ &
		$0.797\pm0.009$ &
        $0.799\pm0.008$ &
		$0.805\pm0.005$ \\ 
		$S_8$ & $0.797\pm0.042$ &
		$0.787\pm0.041$ &
		$0.792\pm0.025$ &
        $0.798\pm0.022$ &
		$0.813\pm0.013$ \\
		\hline
	\end{tabular}
	\caption {Parameter constraints in the $\rm \Lambda$CDM model with $1\sigma$ errors for incremental partial CMB datasets. The dataset in each column includes all of the data listed in the columns to the left of it, so the rightmost column corresponds to the \texttt{CMB} dataset.
    When used without the \texttt{SRoll2} EE data, the \texttt{SPT-3G 2018 TT/TE/EE} dataset includes a prior on $\tau$ in accordance with~\cite{SPT-3G:2022hvq}.}
	\label{tab:tLCDM1}
\end{table}

\begin{table}
	\renewcommand{\arraystretch}{1.2}
    \small
	\centering
	\begin{tabular} {|c|| c |c | c |c|}
		\hline
		& \multicolumn{4}{c|}{$\rm \Lambda$CDM} \\
		\hline
		\hline
		Parameter & CMB & +DESI &  +DES\,Y1 & +Pantheon+ \\
		\hline
		$100\,\omega_b$ & $2.227\pm0.020$ &
		$2.235\pm0.018$ &
		$2.238\pm0.018$ &
		$2.236\pm0.018$  \\ 
		$10\,\omega_{cdm}$ & $1.180\pm0.014$ &
		$1.166\pm0.007$ &
		$1.162\pm0.007$ &
		$1.165\pm0.007$ \\ 
		$H_0$ & $67.80\pm0.65$ &
		$68.46\pm0.31$ &
		$68.60\pm0.30$ &
		$68.49\pm0.30$ \\
		$\tau$ & $0.057\pm0.006$ &
		$0.059\pm0.006$ &
		$0.060\pm0.006$ &
		$0.059\pm0.006$ \\ 
		$\!{\rm ln}(10^{10} A_s)\!$ & $3.045\pm0.012$ &
		$3.048\pm0.012$ &
		$3.047\pm0.011$ &
		$3.046\pm0.011$ \\ 
		$n_s$ & $0.968\pm0.006$ &
		$0.971\pm0.005$ &
		$0.972\pm0.005$ &
		$0.971\pm0.005$ \\ 
		\hline   
		$\rdrag$/\text{Mpc} & $147.75\pm0.37$ &
		$148.04\pm0.27$ &
		$148.10\pm0.27$ &
		$148.06\pm0.27$ \\    
		$\Omega_m$ & $0.307\pm0.009$ &
		$0.298\pm0.004$ &
		$0.296\pm0.004$ &
		$0.297\pm0.004$ \\ 
		$\sigma_8$ & $0.805\pm0.005$ &
		$0.802\pm0.005$ &
		$0.801\pm0.005$ &
		$0.801\pm0.005$ \\ 
		$S_8$ & $0.813\pm0.013$ &
		$0.800\pm0.008$ &
		$0.795\pm0.007$ &
		$0.797\pm0.007$ \\
		\hline
	\end{tabular}
	\caption {Parameter constraints in the $\rm \Lambda$CDM model with $1\sigma$ errors for extended datasets. The dataset in each column includes all of the data listed in the columns to the left of it.}
	\label{tab:tLCDM2}
\end{table}

After constructing the CMB set-up we consider it in combination with DESI BAO data within $\rm \Lambda$CDM. The DESI DR2 results are known to be in moderate tension with full Planck CMB datasets~\cite{DESI:2025zgx}, indicating a possible preference for non-$\rm \Lambda$CDM cosmologies. 
In particular, the inferred values of the matter fraction $\O_m = 0.2977\pm0.0086$ (DESI DR2 with a BBN prior) and $\O_m = 0.3169\pm0.0065$ (Planck PR3) deviate by 3\,$\sigma$. 
This tension is less prominent for our CMB choice, in part because the SPT-Planck combination provides weaker constraints on $\rm \Lambda$CDM parameters, but also because it prefers somewhat higher values of $H_0$ and correspondingly lower values of $\Omega_m$, bringing it within $1.1\sigma$ of the cosmology preferred by DESI DR2 BAO. We note that further addition of DES Y1 weak lensing data and the Pantheon+ catalog does not meaningfully improve constraining power in $\rm \Lambda$CDM due to the relatively weak constraining power of DES Y1 lensing data and the mild tension in $\O_m$ between DESI and Pantheon+. Results of the cosmological parameter inferences for all datasets are presented in Tab.\,\ref{tab:tLCDM2}.

\section{Constraints on neutrino masses $\Mnu$ with SPT-3G 2018}
\label{sec:nuLCDM1}

\begin{figure}[!htb]
    \begin{center}
        \includegraphics[width=1.0\columnwidth]{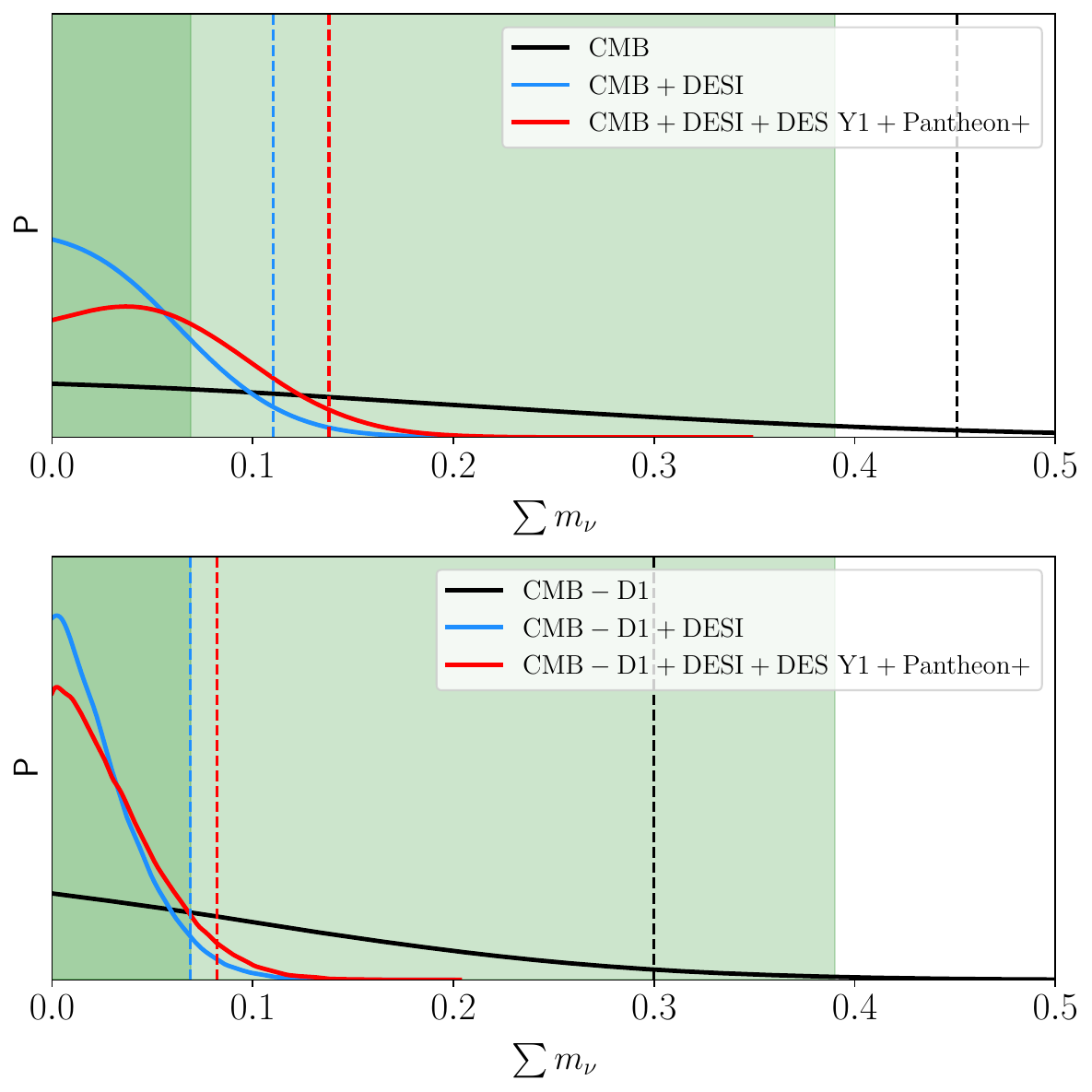}
        \caption {Marginalized 1D posterior distributions (normalized to the unit integral) of $\Mnu$ (in~$\eV$) for datasets based on \texttt{CMB} (upper plot) and \texttt{CMB-D1} (lower plot), with dashed vertical lines marking $95\%$ CL upper limits. Light green shading represents the $95\%$ CL limit from Planck PR4. Dark green shading represents the $95\%$ CL from the DESI DR2 "CMB+BAO" combination.}
        \label{fig:3}
    \end{center}
\end{figure}

Figure\,\ref{fig:3} displays the 2d posterior distributions for different analyses.
The 1D marginalized constraints in the $\rm \Lambda$CDM+$\sum m_\nu$ model are presented in Tab.\,\ref{tab:3}.

Our baseline \texttt{CMB} dataset produces an upper limit of $\Mnu < 0.45 \eV$ at the $95\%$ confidence level, somewhat less strict compared to that produced by the full Planck PR3 analysis ($\Mnu < 0.24 \eV$ for PR3 TT/TE/EE+lensing~\cite{Planck:2018vyg}) but on par with the PR4 limit of $\Mnu < 0.39 \eV$~\cite{Carron:2022eyg}. This can be largely explained by the decreased power of the Planck PR4 lensing compared to PR3 in terms of the effective $\rm A_L$ parameter~\cite{Allali:2024aiv} as well as by the lensing-like feature present in the Planck PR3 $\ell>800$ TT spectrum~\cite{Planck:2016tof}.

The addition of DESI DR2 BAO measurements tightens the constraint to $\Mnu < 0.110 \eV$. This is a less stringent limit than the CMB+BAO results from the DESI DR2 release (e.g. $\Mnu < 0.0691 \eV$ when using the full Planck PR3 TT/TE/EE \texttt{Plik} likelihood and ACT DR6 lensing data~\cite{DESI:2025zgx}) and is in a less tension with ground-based measurements, with the $95\%$ confidence region extending slightly above the $0.1\eV$ experimental lower bound on $\Mnu$ in the case of inverse neutrino hierarchy.

\begin{table}
	\renewcommand{\arraystretch}{1.2}
    \small
	\centering
	\begin{tabular} {|c|| c |c | c |c|}
		\hline
		& \multicolumn{4}{c|}{$\rm \Lambda$CDM+$\sum m_\nu$} \\
		\hline
		\hline
		Parameter & CMB & +DESI & +DES\,Y1 & +Pantheon+ \\
		\hline
        $\Mnu$ & $<0.451$ &
		$<0.110$ &
		$<0.125$ &
        $<0.138$ \\ 
		$100\,\omega_b$ & $2.221\pm0.021$ &
		$2.234\pm0.018$ &
		$2.238\pm0.018$ &
		$2.236\pm0.018$  \\ 
		$10\,\omega_{cdm}$ & $1.192\pm0.021$ &
		$1.168\pm0.008$ &
		$1.163\pm0.008$ &
		$1.165\pm0.008$ \\ 
		$H_0$ & $66.28^{+2.44}_{-1.21}\,(67.05)$ &
		$68.54\pm0.34$ &
		$68.66\pm0.34$ &
		$68.49\pm0.34$ \\
		$\tau$ & $0.057\pm0.006$ &
		$0.059\pm0.006$ &
		$0.060\pm0.006$ &
		$0.059\pm0.006$ \\ 
		$\!{\rm ln}(10^{10} A_s)\!$ & $3.048\pm0.012$ &
		$3.046\pm0.012$ &
		$3.046\pm0.012$ &
		$3.046\pm0.012$ \\ 
		$n_s$ & $0.965\pm0.007$ &
		$0.971\pm0.005$ &
		$0.971\pm0.005$ &
		$0.971\pm0.005$ \\ 
		\hline   
		$\rdrag$/\text{Mpc} & $147.48\pm0.50$ &
		$147.99\pm0.28$ &
		$148.08\pm0.28$ &
		$148.05\pm0.28$ \\    
		$\Omega_m$ & $0.328\pm0.028$ &
		$0.297\pm0.004$ &
		$0.295\pm0.004$ &
		$0.297\pm0.004$ \\ 
		$\sigma_8$ & $0.783\pm0.027$ &
		$0.807\pm0.009$ &
		$0.803\pm0.009$ &
		$0.802\pm0.010$ \\ 
		$S_8$ & $0.817\pm0.015$ &
		$0.803\pm0.010$ &
		$0.797\pm0.009$ &
		$0.798\pm0.010$ \\
		\hline
	\end{tabular}
	\caption {Parameter estimates  in the $\rm \Lambda$CDM+$\sum m_\nu$ model with $1\sigma$ errors for datasets based on SPT-3G 2018 data. The dataset in each column includes all of the data listed in the columns to the left of it. Upper limits on neutrino masses are given at $95\%$ CL. Values in parenthesis represent best-fit points, if the corresponding distributions are noticeably non-gaussian.}
	\label{tab:3}
\end{table}

DES Y1 weak lensing measurements slightly relax the limit to $\Mnu < 0.125\eV$ as DES Y1 prefers a lower range of $S_8$ values~\cite{DES:2017myr}, thus favoring somewhat higher neutrino masses. Including the Pantheon+ SN catalog again loosens the $95\%$ constraint to $\Mnu<0.138 \eV$. Furthermore, the addition of these likelihoods shifts the 1D-marginal mode value of $\Mnu$ from zero to $\sim0.05\eV$ thus indicating a slight preference for massive neutrinos. This is caused by the pull towards lower $H_0$ values when the Pantheon+ SN distance moduli are calibrated by CMB and LSS, resulting in a slight tension between DESI DR2 BAO and Pantheon+~\cite{DESI:2025zgx}.

\begin{table}
	\renewcommand{\arraystretch}{1.2}
    \small
	\centering
	\begin{tabular} {|c|| c |c | c |c|}
		\hline
		& \multicolumn{4}{c|}{$\rm \Lambda$CDM+$\sum m_\nu$} \\
		\hline
		\hline
		Parameter & CMB-D1 & +DESI & +Pantheon+ & +DES Y1 \\
		\hline
        $\Mnu$ & $<0.300$ &
		$<0.069$ &
		$<0.075$ &
        $<0.082$ \\ 
		$100\,\omega_b$ & $2.221\pm0.016$ &
		$2.221\pm0.014$ &
		$2.219\pm0.015$ &
		$2.222\pm0.015$  \\ 
		$10\,\omega_{cdm}$ & $1.204\pm0.016$ &
		$1.179\pm0.007$ &
		$1.181\pm0.007$ &
		$1.176\pm0.007$ \\ 
		$H_0$ & $66.52^{+1.66}_{-0.79}\,(68.41)$ &
		$68.39\pm0.30$ &
		$68.28\pm0.30$ &
		$68.45\pm0.29$ \\
		$\tau$ & $0.057\pm0.006$ &
		$0.062\pm0.006$ &
		$0.061\pm0.006$ &
		$0.061\pm0.006$ \\ 
		$\!{\rm ln}(10^{10} A_s)\!$ & $3.052\pm0.011$ &
		$3.056\pm0.011$ &
		$3.055\pm0.011$ &
		$3.054\pm0.011$ \\ 
		$n_s$ & $0.961\pm0.006$ &
		$0.966\pm0.005$ &
		$0.966\pm0.005$ &
		$0.966\pm0.005$ \\ 
		\hline   
		$\rdrag$/\text{Mpc} & $147.26\pm0.42$ &
		$147.84\pm0.25$ &
		$147.80\pm0.25$ &
		$147.91\pm0.25$ \\    
		$\Omega_m$ & $0.326\pm0.019$ &
		$0.300\pm0.004$ &
		$0.302\pm0.004$ &
		$0.299\pm0.004$ \\ 
		$\sigma_8$ & $0.803\pm0.018$ &
		$0.818\pm0.006$ &
		$0.818\pm0.006$ &
		$0.815\pm0.006$ \\ 
		$S_8$ & $0.835\pm0.012$ &
		$0.818\pm0.008$ &
		$0.820\pm0.008$ &
		$0.813\pm0.007$ \\
		\hline
	\end{tabular}
	\caption {Parameter estimates in the $\rm \Lambda$CDM+$\sum m_\nu$ model with $1\sigma$ errors for datasets based on SPT-3G 2019-2020 (D1+MUSE) data. The dataset in each column includes all of the data listed in the columns to the left of it. Upper limits on neutrino masses are given at $95\%$ CL. Values in parenthesis represent best-fit points, if the corresponding distributions are noticeably non-gaussian.}
	\label{tab:3a}
\end{table}
\section{Comparison to SPT-3G 2019-2020}

The most recent analysis by the SPT-3G collaboration, dubbed D1~\cite{SPT-3G:2025bzu}, covers the 2019 and 2020 observational seasons. Notably, there have been changes made to both the instrument and the analysis pipeline between the seasons of 2018 and 2019, leading to the D1 release not including the 2018 data. Therefore, it is important to track changes in the $\Mnu$ constraints between the 2018 $\texttt{CMB}$ dataset and a version that uses the 2019-2020 SPT-3G data instead. 

For this purpose we construct the \texttt{CMB-D1} dataset, replacing the SPT-3G 2018 $TT/TE/EE$ likelihood with the SPT-3G D1 T$\&$E likelihood~\cite{SPT-3G:2025bzu}. Additionally we update the lensing set to replace the SPT-3G 2018 TT-based lensing spectra with SPT-3G 2019-2020 MUSE EE-based ones \cite{SPT-3G:2024atg}. In both cases we use the primary CMB likelihoods implemented in \texttt{candl}. We keep our cut of the Planck PR3 data unchanged for consistency.

Resulting constraints in the $\Lambda$CDM+$\Mnu$ model are presented in Tab.\,\ref{tab:3a} and in the lower panel of Fig.\,\ref{fig:3}. Compared to \texttt{CMB}, \texttt{CMB-D1} exhibits a $25\%$ reduction in the standard deviations of $\omega_b$ and $\omega_{cdm}$ but a $35\%$ reduction in the uncertainty  of inferred $H_{\rm 0}$ and a similar reduction in the $2\sigma$ upper limit on $\Mnu$. Notable is a $1\sigma$ upward shift in $S_{\rm 8}$, largely due to an increase in $\sigma_8$, indicating a preference for higher matter density contrast from the high-$\ell$ measurements and lensing estimations of SPT-3G 2019-2020. This changes the parameter space preferred in combination with DESI DR2 BAO (the $\sigma_8$ constraint moving up further as $\Omega_m$ is pushed downwards by DESI) towards smaller sum of neutrino masses, $\Mnu<0.07\eV$.
Finally, the addition of the DES Y1 weak lensing and Pantheon+ SN data once again relaxes the constraint to $<0.082\eV$, below the lower mass bound for the inverse hierarchy, and unlike in the SPT-3G 2018 case the marginal distribution of $\Mnu$ still peaks at zero.

\section{Discussion and Conclusions}

We explore a combination of SPT-3G 2018 and Planck CMB spectra as a baseline cosmological dataset leveraging SPT-3G's high angular resolution with minimal reliance on Planck's full CMB maps, reaching an optimal compromise in terms of constraining power. We then utilize this \texttt{CMB} dataset alongside DESI DR2 BAO measurements, DES Y1 weak lensing measurements, and the Pantheon+ SNIa catalog to provide alternative constraints on the sum of active neutrino masses $\Mnu$ in light of the growing tension between cosmological upper limits and oscillatory lower limits on $\Mnu$. We are able to relax the upper limit on $\Mnu$ to $0.110\eV$ for CMB+DESI and to $0.138 \eV$ for CMB+DESI+DESY1+Pantheon+ while remaining competitive in terms of $1\sigma$ errors on $\rm \Lambda$CDM parameters. Additionally, this CMB combination is in better agreement with BBN-calibrated DESI DR2 BAO than both the Planck PR3 and PR4 datasets~\cite{DESI:2025zgx,Allali:2024aiv}.

We then compare this result to the more recent MUSE and D1 data releases from SPT-3G~\cite{SPT-3G:2025bzu} featuring CMB maps obtained during the 2019 and 2020 observational seasons, \emph{but not the 2018 season}. This provides us with an interesting look into the evolution of the $\Mnu$ tension in SPT data within the larger context. In the $\Lambda$CDM+$\Mnu$ model, the most important change between \texttt{CMB} and \texttt{CMB-D1} is the upward shift in $S_8$. While not statistically significant on its own (e.g. the difference in $S_8$ between \texttt{CMB+DESI} and \texttt{CMB-D1+DESI} is at $1.2\sigma$) it pushes the $\Mnu$ posterior against the $0\eV$ prior border, strengthening the tension with ground-based experiments. It has been shown that this behavior represents a more general preference by the DESI data for a lower total matter density than what is inferred from primary and lensing CMB spectra in $\L$CDM~\cite{Cozzumbo:2025ewt,Green:2024xbb,Naredo-Tuero:2024sgf}. Indeed, the \texttt{SPT-3G D1} dataset produces a $\L$CDM constraint of $\O_m = 0.325\pm 0.009$~\cite{SPT-3G:2025bzu}, 1.4$\s$ above our \texttt{CMB} result. This shift shrinks the allowed mass range once $\Mnu$ is free to vary and enhances the push toward zero (or formally negative) neutrino mass exhibited by DESI BAO data, ensuring that even with the addition of the DES Y1 and Pantheon+ likelihoods the marginalized $\Mnu$ posterior peak is locked at zero.

The deficit of matter density in the DESI analysis compared to predictions from CMB ties the $\Mnu$ tension to both the 'Hubble crisis' and the complicated discussion around $S_8$ measurements from different lensing and clustering signals~\cite{CosmoVerseNetwork:2025alb,McCarthy:2024tvp,Artis:2024zag}. This work demonstrates that while the DESI measurements are now the main driver of this tension (and so are being actively probed for potential contradictions or a theoretical explanation~\cite{CosmoVerseNetwork:2025alb}) the choice of a CMB baseline does still affect the $\Mnu$ constraint. In the case of SPT-3G, there have been changes both in hardware and in the analysis pipeline between the 2018 and 2019-2020 observational runs. It is important to fully understand how these changes shape the resulting predictions beyond $\L$CDM. It would be interesting to see, for example, what constraints are produced by re-analyzing the 2018 raw data using the D1 pipeline. We leave such exploration to future work.

\vspace{1cm}
\section*{Acknowledgments}

This work is supported in the framework of the State project ``Science'' by the Ministry of Science and Higher Education of the Russian Federation under the contract 075-15-2024-541. All numerical calculations have been performed on the HybriLIT heterogeneous computing platform (LIT, JINR) (\href{http://hlit.jinr.ru}{http://hlit.jinr.ru})  


\bibliographystyle{JHEP}
\bibliography{short.bib}

\end{document}